\documentclass[12pt]{article}

\usepackage{epsfig}

\newcommand{\eq}{\begin{equation}}
\newcommand{\eqx}{\end{equation}}
\newcommand{\eqn}{\begin{eqnarray}}
\newcommand{\eqnx}{\end{eqnarray}}

\newcommand{\nc}{$ N_{\rm cut}\  $}

\newcommand\txf[2]{{\textstyle{#1\over#2}}}
\newcommand\half{\txf12}
\newcommand\fourth{\txf14}

\newcommand\erfc{\mathop{\rm erfc}\nolimits}

\begin{document}

\date{September 17, 2002}

\title{
\null\vskip -60 pt
\begin{flushright}
\normalsize\rm TPJU-15/02, IFUP-TH/2002-36, MPI-PhT/2002-45
\end{flushright}
\vskip 10 pt
Exact Witten Index in D=2 supersymmetric Yang-Mills quantum mechanics  }

\author{M. Campostrini,\\
\normalsize \it INFN, Sezione di Pisa, and\\
\normalsize \it Dipartimento di Fisica ``Enrico Fermi'' dell'Universit\`a di Pisa,\\
\normalsize \it Via Buonarroti 2, I-56125 Pisa, Italy,\\
\\J. Wosiek,\\
\normalsize \it M. Smoluchowski Institute of Physics,
Jagellonian University,\\
\normalsize \it Reymonta 4, 30-059 Krak\'{o}w, Poland\\
}

\maketitle

\begin{abstract}
A new, recursive  method of calculating matrix elements of polynomial hamiltonians is 
proposed. It is particularly suitable for the recent algebraic studies of the supersymmetric
Yang-Mills quantum mechanics in any dimensions. For the $D=2$ system with the SU(2)
gauge group, considered here, the technique gives exact, closed expressions for arbitrary matrix 
elements of the hamiltonian and of the supersymmetric charge, in the occupation number representation.
Subsequent numerical diagonalization provides  spectrum and restricted Witten index of the system 
with very high precision (taking into account up to $10^5$ quanta).

Independently, the exact value of the restricted Witten index is derived analytically for the first time.  
\end{abstract}

\section{Introduction}
Supersymmetric quantum mechanics provides a simple laboratory to study many properties of 
supersymmetric systems \cite{WI,CH,CO}. Recently
supersymmetric Yang-Mills quantum mechanics (SYMQM) in $d$ space dimensions attracts
a lot of attention because 
of its possible relation with M-theory \cite{BFSS} for $d=9$ and large number of colours.
Even though the three loop calculations \cite{DEG,WT}
question the exact equivalence between the two, it still remains a valuable model of the latter sharing many of its
features (e.g., continuum spectrum of scattering states with the threshold bound state -- a supergraviton). 
At the same time, studies of the lower-dimensional systems with various gauge 
groups provide the global understanding
of the whole family of these models with many interesting limiting cases. A good example is the $D=4$ SYMQM with the
SU(2) gauge group whose spectrum, in the zero fermion sector, is identical with that of the well known 0-volume 
glueballs \cite{L,LM,VABA,JW}. Moreover, supersymmetry guarantees that, in addition 
to the continuum of the scattering states, 
there must also exist localized gluino-glueball bound states with the same masses. Indeed such solutions were found
in Ref.\ \cite{JW} where a new algebraic approach to study these systems was proposed. Compact version of the $D=4$
SYMQM was also studied recently by van Baal who pushed rather far analytical 
understanding of this system \cite{vBS,vBN}.

The technique of \cite{JW} is the adaptation of the hamiltonian methods \cite{CMB1} to supersymmetric systems
with local gauge invariance. 
It consists of two steps: first, a finite (e.g., cut off) basis of gauge-invariant eigenstates 
of the bosonic and fermionic occupation number operators $\hat{B}=a_b a^{\dagger}_b$ and $\hat{F}=f_b f^{\dagger}_b$
is generated (we omit all but colour indices for a moment). Second, the matrix representation of the hamiltonian
(and any other relevant observable) is derived and numerically diagonalized. All this is automated by implementing 
standard rules of quantum mechanics in an algebraic language like Mathematica. A faster, compiler based, version is
now available \cite{JK}. As the cutoff \nc we choose the number of bosons $B$, i.e., the cut off basis
consists of all states with $B \le N_{\rm cut} $ and all allowed fermions. The cutoff is gauge and rotationally invariant 
and consequently the spectrum reveals the full SO($D{-}1$) symmetry for each $N_{\rm cut}$.
The technique has been applied to Wess-Zumino quantum mechanics, $D=2$
and $D=4$ SYMQM and to $D=5$--10 YMQM, all based on the SU(2) gauge group \cite{JW,JW2}.  
In all cases studied until now the spectrum of lower states converges with \nc before the number of states 
becomes unmanageable. Calculations become more and more time consuming with increasing 
$D$ and $N$ but the $D=10$ case
is within the reach of today's computers for the first few values of $N$. Similar methods have been independently
developed in Refs \cite{JAP,P1} to study lower dimensional supersymmetric field theories.
 
In the first part of this letter we present a new method to calculate matrix representations
 and apply it to the $D=2$ SYMQM in Section 3. 
We calculate recursively the matrix elements of the hamiltonian thereby eliminating  
lengthy and space consuming process of generation and storing of the basis. 
For the $D=2$ system recursions 
can be solved resulting in closed expressions for any matrix element of the hamiltonian $H$. 
We then diagonalize $H$ numerically
and calculate the restricted Witten index for this system to virtually arbitrary precision. 
The method can be generalized 
to higher $D$, allowing to reach values of the cutoff larger than those in Ref.\ \cite{JW} .

The second part contains an exact calculation of the restricted index, exploiting the analytic properties of the 
densities of bosonic and fermionic states, suitably regularized in the infrared. To our knowledge this is the first 
analytic calculation of the restricted index for this system.

\section{Supersymmetric Yang-Mills quantum mechanics in two dimensions}

This model is exactly soluble \cite{CH} also for higher gauge groups \cite{STS}.
It is however still interesting as it shares some of the complexity of higher dimensional models. 
For example it has a continuous spectrum which is a characteristic feature of many other supersymmetric models
and poses some challenge for the hamiltonian methods. Moreover it has a nontrivial Witten index 
which was defined only recently \cite{JW}.

The system, reduced from $D=2$ to one (time) dimension \cite{BRINK}, is described by the
three real bosonic variables $x_a(t)$ and three complex, fermionic degrees
of freedom $\psi_a(t)$, both in the adjoint representation of SU(2) , $a=1,2,3$.

The hamiltonian reads \cite{CH}
\eq
H={1\over 2} p_a p_a +i g \epsilon_{abc} \psi_a^{\dagger} x_b \psi_c, \label{HD2}
\eqx
where the quantum operators $x,p,\psi,\psi^{\dagger}$ satisfy the canonical (anti)commutation rules
\eq
[x_a,p_b]=i\delta_{ab} ,\;\; \{\psi_a,\psi_b^{\dagger}\}=\delta_{ab},
\eqx
and can be written in terms of the creation and annihilation operators 
\eqn
   x_a={1\over\sqrt{2}}(a_a+a_a^{\dagger}), & p_a={1\over i \sqrt{2}}(a_a-a_a^{\dagger}),  \\ \label{XPD2}
   \psi_a=f_a, & \psi_a^{\dagger}=f_a^{\dagger}.
\label{eq:f}
\eqnx
The system has a gauge invariance with generators
\eq
G_a=\epsilon_{abc}(x_b p_c - i \psi_b^{\dagger} \psi_c). \label{GD2}
\eqx
Therefore the physical Hilbert space consists only of the gauge-invariant states.
This constraint is easily accommodated by constructing all possible combinations
of creation operators (creators) invariant under SU(2), and using them to generate a complete
gauge-invariant basis of states. There are four lower order creators:
\eq
(aa)  \equiv  a_a^{\dagger} a_a^{\dagger},\;\; 
(af)  \equiv  a_a^{\dagger} f_a^{\dagger},\;\;
(aff) \equiv  \epsilon_{abc}a_a^{\dagger} f_b^{\dagger}f_c^{\dagger},\;\;
(fff) \equiv \epsilon_{abc}f_a^{\dagger} f_b^{\dagger}f_c^{\dagger}. \label{creators} 
\eqx
Fermionic creators satisfy $ (ff)=(af)^2=(aff)^2=(fff)^2=0 $, therefore the whole basis can be conveniently
organized into the four towers of states, each tower beginning with one of the following states
\eq
|0_F\rangle=|0\rangle;\;\; |1_F\rangle=(af)|0\rangle,\;\; |2_F\rangle=(aff)|0\rangle,\;\; |3_F\rangle=(fff)|0\rangle, 
\label{ground}
\eqx
where we have labeled the states by the gauge-invariant fermionic number $F=f_a f_a^{\dagger}$.
To obtain the whole basis
it is now sufficient to repeatedly act on the four vectors (\ref{ground}) with the bosonic creator $(aa)$.  
Acting with other creators either gives zero, due to the Pauli principle, or produces a state
from another tower, already obtained by application of $(aa)$. 
The basis with cutoff \nc is then obtained by applying $(aa)$ up to \nc times to each of the four
``base'' states of Eq.\ (\ref{ground}). Obviously our cutoff is 
gauge-invariant, since it is defined in terms of the gauge-invariant creators.

The hamiltonian (\ref{HD2}) reduces in the physical basis to that of a free bosonic particle 
\eq
H={1\over 2} p_a p_a + g x_a G_a, \label{HD2free}
\eqx
therefore it preserves the fermionic number and can be diagonalized independently
in each sector spanned by the four towers in Eq.~(\ref{ground}).

The spectrum is doubly degenerate even at finite $N_{\rm cut}$, because of the particle-hole
symmetry which is preserved by our cutoff.  Particle-hole symmetry
relates empty and filled fermionic states ($ |0_F\rangle
\leftrightarrow |3_F\rangle $) and their 1-particle 1-hole
counterparts ($ |1_F\rangle \leftrightarrow |2_F\rangle $). 
On the other hand, the supersymmetry generator
\begin{equation}
Q = \sum_a \psi_a p_a ,
\label{eq:Q}
\end{equation}
connects sectors which differ by 1 in the fermionic number, 
e.g., it connects $0_F - 1_F$ sectors and $2_F - 3_F$ sectors, but it {\em does not} connect
$1_F$ and $2_F$ sectors. This  can be easily seen in terms of the ``$B$-parity'', $P_B=(-1)^B$: 
$Q$ not only changes $F$, but also $P_B$; however, the $F=1$ and $F=2$ sectors have the same $P_B$,
hence matrix elements of $Q$ between these sectors vanish.

\noindent{\em Witten index vs restricted index.}
Because the particle-hole symmetry
interchanges odd and even fermionic numbers, the Witten index 
\eq
I_W(T)=\Sigma_{i} (-1)^{F_i}\exp{(-T E_i)},
\eqx
vanishes identically for this model\footnote{This is also true for any finite cutoff since \nc
preserves the particle-hole symmetry.}. Nevertheless one can obtain a nontrivial and interesting 
information by defining the index {\em restricted} to a one pair (e.g., $0_F - 1_F$) of sectors
\eq
I(T)^{(0,1)}=\Sigma_{i,F_i=0,1} (-1)^{F_i}\exp{(-T E_i)}.
\eqx
Since $Q$ does not connect the $1_F$ sector with the $2_F$ sector,
supersymmetry balances independently, and identically, fermionic and bosonic states within 
the $0_F - 1_F$ and $2_F - 3_F$ pairs, with the usual exception of the vacuum.
Therefore the restricted index is a good and nontrivial measure of the amount
of the violation of SUSY, even when the total Witten index
vanishes.  Obviously, $I^{(2,3)}=-I^{(0,1)}$.
In principle, one could also consider $I^{(0,3)}$ and $I^{(1,2)}$,
however, similarly to the global index, they vanish due to the particle-hole symmetry.
Studying the restricted
index is particularly interesting in this model since, due to the
continuum spectrum, it does not have to be an integer.

\section{Exact  matrix elements and numerical diagonalization}

In order to simplify numerical computations, it is useful to avoid
dealing explicitly with the Fock space vectors. This can be achieved by writing every quantity of
interest as a vacuum expectation value (vev) of a gauge-invariant
operator, and deriving recursive relations between such vev's.

We shall deal here explicitly with the $F=0$ and $F=1$ sectors, the $F=2$ and $F=3$
sectors can be obtained by the particle-hole symmetry. To begin with,
let us introduce the gauge-invariant operators
\begin{equation}
A = \sum_b a_b a_b, \qquad 
A^\dagger = \sum_b a^\dagger_b a^\dagger_b = (aa), \qquad
B = B^\dagger = \sum_b a_b a^\dagger_b,
\end{equation}
which satisfy the commutation relations
\begin{equation}
[A,A^\dagger] = 4 B - 6, \qquad
[A,B] = 2 A.
\label{eq:ABcomm}
\end{equation}

\noindent States in the $F=0$ sector have the form
\begin{equation}
|2n,0\rangle = {1\over\sqrt{c_n}} (A^\dagger)^n|0\rangle,
\end{equation}
and those with $F=1$  can be written as
\begin{equation}
|2n{+}1,1\rangle = {1\over\sqrt{c'_n}} F^\dagger (A^\dagger)^n|0\rangle,\qquad
F = \sum_b a_b f_b, \quad
F^\dagger = \sum_b a^\dagger_b f^\dagger_b = (af).
\end{equation}
Observe that in the $F=0$ sector $f_a f^\dagger_b=\delta_{ab}$
and therefore $F F^\dagger = B$.
We are interested in the scalar products
\begin{eqnarray}
\langle0|A^{n'} (A^\dagger)^n|0\rangle &=& \delta_{n'n} c_n,
\label{eq:vev1} \\
\langle0|A^{n'}F F^\dagger(A^\dagger)^n|0\rangle &=& 
\langle0|A^{n'} B (A^\dagger)^n|0\rangle = \delta_{n'n} c'_n,
\label{eq:vevB}
\end{eqnarray}
and in the matrix elements of $H$ and $Q$. Since
\begin{equation}
H = -\fourth(A+A^\dagger - 2B + 3),
\end{equation}
the matrix elements of $H$ can be written in terms of the above
vev's and of $\langle0|A^{n'} B^2 (A^\dagger)^n|0\rangle$. Given Eqs.\
(\ref{eq:f}) and (\ref{eq:Q}), the matrix elements
of $Q$ are also expressible by the vev's defined in Eqs.\ 
(\ref{eq:vev1}) and (\ref{eq:vevB}).

A general technique to compute the desired vev's  
 is to consider a generic matrix element
\begin{eqnarray}
&&\langle0|A^{n_1}\Theta A^{n_2}\Xi(A^\dagger)^n|0\rangle \nonumber \\
&&\quad=\, \langle0|A^{n_1-1}\Theta A^{n_2+1}\Xi(A^\dagger)^n|0\rangle
+ \langle0|A^{n_1-1}[A,\Theta]A^{n_2}\Xi(A^\dagger)^n|0\rangle,
\label{eq:recursion}
\end{eqnarray}
where $\Theta=B,A^\dagger$ and $\Xi=1,B$. Eq.\ (\ref{eq:recursion})
allows to ``shift to the left'' $\Theta$ until it is immediately to
the right of $\langle0|$. Then we use $\langle0|A^\dagger=0$ or
$\langle0|B=\langle0|3$; the commutator terms are of lower order and
therefore the iteration of Eq.\ (\ref{eq:recursion}) closes,
giving the desired recursion. 

This method can be applied to SYMQM in arbitrary dimension. For $D=2$,
recursions can be solved, providing explicit expressions for the
matrix elements of $H$ and $Q$ between orthonormalized states.
To this end we consider first the $F=0$ sector and define
\begin{equation}
\langle 0|A^n A^\dagger = l_n \langle 0|A^{n-1},
\end{equation}
$l_n$ can be computed recursively: $l_1=6$ and
\begin{eqnarray}
l_n \langle 0|A^{n-1} &=& \langle 0|A^{n-1} AA^\dagger  =
\langle 0|A^{n-1}(A^\dagger A + 4B - 6) \nonumber \\
&=&
\langle 0|[(l_{n-1} - 6)A^{n-1} + 4([A^{n-1},B]+BA^{n-1})] =
(l_{n-1} + 8n - 2)\langle 0|  \nonumber \\
\end{eqnarray}
therefore $l_n = 2n + 4n^2$.  Moreover, $c_n = l_n\,c_{n-1}$ and
$c_0=1$.  
We next compute 
\begin{equation}
\langle 2n,0|B|2n,0\rangle = 
{1\over c_n} \langle0|A^n B (A^\dagger)^n|0\rangle =
{1\over c_n} \langle0|(2n+B)A^n (A^\dagger)^n|0\rangle = 2n+3
\end{equation}
and
\begin{equation}
\langle 2n{-}2,0|A|2n,0\rangle = \langle 2n,0|A^\dagger|2n{-}2,0\rangle 
 = \sqrt{c_n\over c_{n-1}} = \sqrt{2n+4n^2}.
\end{equation}
Therefore, the nonzero matrix elements of $H$ are
\begin{eqnarray}
&&\langle 2n,0|H|2n{-}2,0\rangle = \langle 2n{-}2,0|H|2n,0\rangle 
= -\fourth\sqrt{2n+4n^2}, \\
&&\langle 2n,0|H|2n,0\rangle =  n + \txf34.  \label{nHn}
\end{eqnarray}

The $F=1$ sector is dealt with in the same way, and the result is
\begin{eqnarray}
&&\langle 2n{+}1,1|H|2n{-}1,1\rangle = \langle 2n{-}1,1|H|2n{+}1,1\rangle 
   = -\txf14\sqrt{6n+4n^2} \\
&&\langle 2n{+}1,1|H|2n{+}1,1\rangle = n + \txf54.
\end{eqnarray}
For the supersymmetry generator we obtain
\begin{equation}
\langle 2m,0|Q|2n{+}1,1\rangle = 
-i\sqrt{m+\txf32}\,\delta_{m,n} + i\sqrt{m}\,\delta_{m,n+1}.
\end{equation}

In the cut Hilbert space, with  states up to the $|2N_{\rm
cut}{+}2,0\rangle$ and $|2N_{\rm cut}{+}1,1\rangle$, 
$Q$ is represented by the $(N_{\rm cut}{+}2){\times}(N_{\rm cut}{+}1)$ matrix
$Q_{ij}=\langle 2i,0|Q|2j{+}1,1\rangle$ and $Q^\dagger$ by its adjoint.
Define
\begin{equation}
\langle 2i,0|H_{\rm cut}|2j,0\rangle =
    \half\sum_k Q_{ik} Q^\dagger_{kj}, \quad
\langle 2i{+}1,1|H_{\rm cut}|2j{+}1,1\rangle =
    \half\sum_k Q^\dagger_{ik} Q_{kj}.
\end{equation}
All matrix elements of $H$ and $H_{\rm cut}$ are equal, apart from 
\begin{equation}
\langle 2N_{\rm cut}{+}2,0|H_{\rm cut}|2N_{\rm cut}{+}2,0\rangle =
\half(N_{\rm cut}+1),
\end{equation}
cf.\ Eq.(\ref{nHn}). 
Hence we confirm that we can define a cut off hamiltonian $H_{\rm cut}$
whose spectrum  has exact supersymmetry at any finite \nc \cite{JW,JAP,P1}.

The hamiltonian matrix has a tridiagonal structure and it can be
diagonalized numerically in a very efficient way. We use the $O(N^2)$
algorithm implemented in the lapack library, which computes all
eigenvalues for $N_{\rm cut}=10^5$ in a few minutes on a PC.

\begin{figure}[tb]
\centerline{\psfig{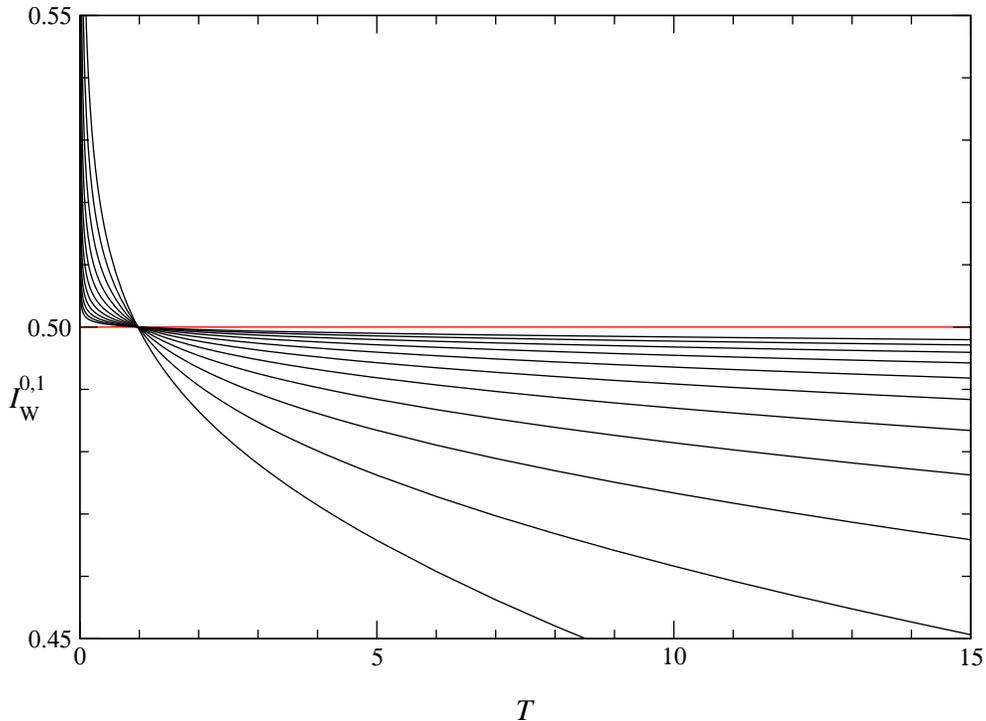}}
\caption{The restricted Witten index computed for values of 
$N_{\rm cut}$ ranging from 125 to 128000.}
\label{fig:Iw1}
\end{figure}

A plot of the restricted index as a function of the Euclidean time
is presented in Fig.\ \ref{fig:Iw1} confirming results of \cite{JW} to a much higher precision,
and strongly suggesting the exact, time independent, value $I(T)=1/2$.

Note the intriguing multiple crossing at $T=1$ where the index seems to
attain its asymptotic value for all \nc . Upon closer inspection it turns out that
the curves do not cross at the same point. However the intersection
points approaches $I_W=1/2$, $T=1$ as $\delta I_W=O(1/N_{\rm cut})$, $\delta T=O(1/\sqrt{N_{\rm cut}})$.
This suggests existence of a ``duality'' transformation which relates, for finite $N_{\rm cut}$,
SUSY violations above and below the ``critical'' energy $E \sim 1$.

\section{Exact calculations of the restricted index}
Gauge-invariant eigenstates of the free hamiltonian, Eq.~(\ref{HD2free})
are labeled by the absolute value of the momentum $p$. In this representation the index reads%
\footnote{We consider the index restricted to the $F=0$ and $F=1$ sectors but will omit 
for the simplicity the $(0,1)$ superscript.}
\eq
I(T)=\int_0^{\infty} \left( \rho_B(p) - \rho_F(p) \right) e^{- T p^2/2} dp,
\eqx 
    
Since the spectrum is continuous, definition of the bosonic and fermionic densities, 
$\rho_B$ and $\rho_F$, requires
an infrared regularization. 
Therefore we first solve the free problem in a spherical well of radius $R$ and then 
calculate the index as a limit\footnote{We thank Ken Konishi for the discussion on this point.}. 
\eq
I(T)=\lim_{R\rightarrow\infty} I_R(T),
\eqx
with the IR regularized index given by the discrete sum
\eq
I_R(T)=\sum_n \left( e^{-T(p^B_n)^2/2} - e^{-T(p^F_n)^2/2} \right),  \label{discr}
\eqx
where $p^B_n (p^F_n)$ are the discrete momenta of the bosonic and fermionic states respectively.

The spectrum of a free particle in a three dimensional spherical well of radius $R$ is determined by the
boundary condition
$
j_l(pR)=0,
$
for the $l$-th spherical wave. However the local gauge symmetry 
limits allowed values of $l$  to $l=0,1$. Only these two angular momenta can be combined
with the fermionic colour spin, $s=1$, into  a scalar, $J=0$, gauge-invariant state in the $F=0$ and $F=1$ 
sectors.\footnote{The ``angular momentum'' considered here is the colour angular momentum
 which generates rotations in the three dimensional colour space.}
It follows that the sum in Eq.~(\ref{discr}) is over the positive zeroes $z^{(l)}_n$ of the
first two Bessel functions. 
\eq
p^B_n=z^{(0)}_n/R,\;\;\; p^F_n=z^{(1)}_n/R,
\eqx

A simple way to compute the $R\to\infty$ limit of Eq.~(\ref{discr}) is to
replace $z^{(1)}_n$ by its large-$n$ asymptotic expansion
\begin{equation}
z^{(0)}_n = \pi n, \qquad
z^{(1)}_n = \beta_n - 1/\beta_n + O(n^{-3}), \qquad
\beta_n=\pi(n+\half);
\end{equation}
we note that that, in the $R\to\infty$ limit, higher-order terms
do not contribute and the sum can be replaced by an integral:
\begin{eqnarray}
I_{\rm W}(T) &=& \lim_{u\to0} \int_0^\infty dn \left\{
   \exp\left[-u^2 n^2\right] - \exp\left[-{2u^2\over\pi^2}\right]
   \exp\left[-u^2(n+\half)^2\right]\right\} \nonumber \\
&=& \lim_{u\to0}{\sqrt{\pi}\over2u} \left\{1 -
        \exp\left[-{2u^2\over\pi^2}\right]
        \erfc\left[u\over2\right]\right\}
 = {1\over2},
\end{eqnarray}
where 
\begin{equation}
u^2 = {\pi^2 T\over2R^2},\qquad
\erfc(z) = {2\over\sqrt{\pi}}\int_z^\infty\exp[-t^2]\,dt,\qquad
\erfc(0) = 1.
\end{equation}

A more illuminating strategy to compute $I_{\rm W}$ starts with
 the well known theorem on meromorphic functions to rewrite
Eq.~(\ref{discr}) as a contour integral
\eq
I_R(T)=\frac{1}{2\pi i} \int_\Gamma  e^{-T (z/R)^2/2} h(z) dz, \label{cont}
\eqx 
where $\Gamma$ denotes any contour enclosing counterclockwise positive real axis 
and contained within 
the angular region $ -\pi/4 < \arg(z) < \pi /4 $.  The weight $h$ is given by
\eq
h(z)=j_0'(z)/j_0(z) - j_1'(z)/j_1(z) +1/z. \label{h}
\eqx
The first two terms follow directly from the above theorem, the last one subtracts the  
pole at $z=0$. Since $j_1(z)$ has a second-order zero
at $z=0$ the theorem would not apply. However we can subtract by hand the singular 
terms of the Laurent expansion around $z=0$ from both $j_0$ and $j_1$,
since the $z_0=0$ contributions cancel in Eq.~(\ref{discr}) anyway. 

Now, choose for $\Gamma$ two straight lines%
\footnote{Contribution from the section of the big circle is negligible at large $R$.} 
at angles $\pm \vartheta$, $0 < \vartheta < \pi/4 $, running to (from) the origin from (to) infinity, 
and rescale $z$ . This gives
\eq
I_R(T)=\frac{R}{2\pi i} \int_\Gamma  e^{-T u^2/2} h(u R) du, \label{contt}
\eqx
with the same contour $\Gamma$ in the $u$ plane.

The whole $R$ dependence of the integrand is now in $h(u R)$. When $R\rightarrow \infty$, poles on the real axis 
condense into a cut. However above and below the cut the integrand simplifies considerably for large $R$. We exploit
this by deforming the contour into two lines parallel to the real axis at distance $\epsilon$ 
 and a vertical section along the imaginary axis.
Due to the Schwarz reflection principle, contribution from the vertical section vanish. Hence $I_R(T)=I_R^+ + I_R^-$ with
\eq
I_R^{\pm}(T) = \mp R \frac{1}{2\pi i} \int_0^{\infty}  e^{-T (p\pm i\epsilon)^2/2} h(R(p\pm i\epsilon)) dp, 
\label{con}
\eqx
At large $R$ and {\em fixed} $\epsilon$ trigonometric functions are dominated by the terms
growing exponentially with $R$
and in fact cancel in the ratios in Eq.~(\ref{h}). Consequently we have simply (e.g., above the cut)
\eq
\lim_{R\rightarrow\infty,\; \epsilon > 0\; \rm fixed} h(R(p + i\epsilon))= \frac{z-2 i}{z(z-i)}, \;\;\; 
z=R(p + i\epsilon).
\eqx
and symmetrically below the cut.

Now we can move toward the upper and lower ends of the cut by taking $\epsilon\rightarrow 0$. 
The 
imaginary contributions from upper and lower contours cancel and our main result reads
\eq
I(T)=\lim_{R\rightarrow\infty}\frac{R}{\pi}\int_0^{\infty} e^{-T p^2/2} \frac{dp}{R^2 p^2 +1}. \label{main}
\eqx
This can be easily calculated with the aid of the proper time integral representation
\eqn
&&I(T)=\lim_{R\rightarrow\infty}\frac{R}{\pi}\int_0^{\infty} e^{-T p^2/2} \int_0^{\infty} e^{-s(R^2p^2+1)} ds dp  \\
  &&=\lim_{R\rightarrow\infty}\frac{R}{\sqrt{2\pi}}\int_0^{\infty}  
  \frac{e^{-s} ds}{\sqrt{2 R^2 s +T}}=\frac{1}{2\sqrt{\pi}}\int_0^\infty e^{-s}\frac{ds}{\sqrt{s}} =
   \frac{1}{2}, \nonumber
\eqnx
which finally proves our conjecture based on Fig.\ \ref{fig:Iw1}
The limit (\ref{main}) singles out only the contribution from the $p=0$ state. In fact  Eq.~(\ref{main}) 
is nothing but
\eq
I(T)=\int_0^{\infty} \delta(p) e^{-T p^2/2} dp . 
\eqx
This form clearly proves that supersymmetry is restored in the $R\rightarrow\infty$ limit.
 Densities of the bosonic
and fermionic states are exactly equal for any nonzero energy, while the lowest state does 
not have a supersymmetric partner, which results in the point-like contribution from the zero momentum. 
This also confirms our earlier
assertion that, in the $D=2$ system, restricted Witten index is time independent even 
for the continuum spectrum \cite{ST}.

Summarizing, supersymmetric quantum mechanical systems can now be solved with better
precision in any dimensions. This provides a useful numerical tool for the nonperturbative 
solutions of matrix theories. Here it was applied to the $N=2$ case. However, since the
present recursive approach uses only gauge-invariant operators, it is even more useful
for higher gauge groups. Finally, analytical calculation of the restricted Witten index
for the $D=2$ showed explicitly how the supersymmetry is restored while removing
the infrared cutoff.

\noindent {\em Acknowledgments.} 
We thank Ken Konishi and Pierre van Baal for
intensive and very useful discussions. J.W. thanks the Theory Group at LSU,
 especially Richard W. Haymaker, for their hospitality. J. W. also thanks 
 the Theory Group of the Max-Planck-Institute in Munich for their hospitality.
 This work is  supported by the
Polish Committee for Scientific Research under the grant no. PB 2P03B09622,
during 2002 -2004.


\begin{thebibliography}{99}

\bibitem{WI} E. Witten, Nucl. Phys. {\bf B185/188} (1981) 513.
\bibitem{CH} M. Claudson and M. B. Halpern, Nucl. Phys. {\bf B250} (1985) 689.
\bibitem{CO} F. Cooper, A. Khare and U. Sukhatme, Phys. Rep.
 {\bf 251}(1995) 267, hep-th/9405029.
\bibitem{BFSS} T. Banks, W. Fishler, S. Shenker and L. Susskind,
 Phys.Rev. {\bf D55} (1997) 6189; hep-th/9610043.
\bibitem{DEG} M. Dine, R. Echols and J. P. Gray, Nucl. Phys. {\bf B564} (2000) 225,
hep-th/9810021.
\bibitem{WT} W. Taylor, Rev. Mod. Phys. {\bf 73} (2001) 419, hep-th/0101126.
\bibitem{L} M. L\"{u}scher, Nucl. Phys. {\bf B219} (1983) 233.
\bibitem{LM} M. L\"{u}scher and G. M\"{u}nster, Nucl. Phys. {\bf B232} (1984) 445.
\bibitem{VABA} P. van Baal, Acta Phys. Polon. {\bf B20} (1989) 295.
\bibitem{JW} J. Wosiek, Nucl. Phys. {\bf B} {\em in print}, hep-th/0203116.
\bibitem{vBS} P. van Baal, in {\em At the Frontiers of Particle Physics - Handbook of QCD, 
Boris Ioffe Festschrift},
vol. 2, ed. M. Shifman (World Scientific, Singapore 2001) p.683; hep-ph/0008206.
\bibitem{vBN} P. van Baal, hep-th/0112072, to appaear in the Michael Marinov Memorial
 Volume {\em Multiple Faces of Quantization and Supersymmetry}, ed. M. Olshanetsky and
 A. Vainstein, World Scientific.
\bibitem{CMB1} C. M. Bender et al., Phys. Rev. {\bf D32} (1985) 1476.
\bibitem{JK} J. Kotanski and J. Wosiek, to appear in the Proceedings of the XX Symposium 
on Lattice Field Theory, MIT, Cambridge MA, June 2002, hep-lat/0208067. 
\bibitem{JW2} J. Wosiek, to appear in the Proceedings of the NATO Workshop on QCD, Stara Lesna, 
Slovakia, January 2002, hep-th/0204243.
\bibitem{JAP}  Y. Matsumura, N. Sakai and T. Sakai, Phys. Rev. {\bf D52} (1995) 2446.
\bibitem{P1} J. R. Hiller, S. Pinsky and U. Trittmann, hep-th/0112151, hep-th/0106193.
\bibitem{STS} S. Samuel, Phys. Lett. {\bf B411} (1997) 268, hep-th/9705167.
\bibitem{BRINK} L. Brink, J. H. Schwarz and J. Scherk, Nucl. Phys. {\bf B121} (1977) 77.
\bibitem{ST} S. Sethi and M. Stern, Comm. Math. Phys., {\bf 194} (1998) 675, hep-th/9705046.

\end{thebibliography}
\end{document}